\documentclass[envcountsect,envcountsame]{llncs}
\usepackage{makeidx}
\usepackage{amsfonts}
\usepackage{amsmath}
\usepackage{amssymb}
\usepackage{amscd}
\usepackage[dvips]{graphicx}
\usepackage{algorithmic}
\usepackage{algorithm}
\usepackage{comment}
\usepackage{color}
\usepackage{array,amssymb,amsmath,amsbsy}
\usepackage{enumerate}
\usepackage[bottom]{footmisc}
\usepackage[all,cmtip]{xy}
\usepackage{amsbsy}
\usepackage{a4wide}
\usepackage{hyperref}
\usepackage[a4paper,margin=0.95in]{geometry}
\usepackage{xspace}
\usepackage{lineno}
\let\doendproof\endproof
\renewcommand\endproof{~\hfill$\qed$\doendproof}
\makeatletter
\renewcommand*\l@author[2]{}
\renewcommand*\l@title[2]{}
\makeatletter
\newcommand{\nocontentsline}[3]{}
\newcommand{\tocless}[2]{\bgroup\let\addcontentsline=\nocontentsline#1{#2}\egroup}

\newenvironment{packed_enum}{
    \begin{enumerate}
        \setlength{\itemsep}{1pt}
        \setlength{\parskip}{0pt}
        \setlength{\parsep}{0pt}
}{\end{enumerate}}

\newenvironment{packed_itemize}{
    \begin{itemize}
        \setlength{\itemsep}{1pt}
        \setlength{\parskip}{0pt}
        \setlength{\parsep}{0pt}
}{\end{itemize}}

\newcommand{\heading}[1]{\smallskip\par\noindent{\bfseries #1}}

\def\computationproblem#1#2#3#4{
	\vskip 1ex
	\begin{center}
	\fbox{\begin{tabular}{rp{#4}}
	{\bfseries Problem:}\enspace&#1\\
	{\bfseries Input:}\enspace&#2\\
	{\bfseries Question:}\enspace&#3\\
	\end{tabular}}
	\end{center}
	\vskip 1ex
}

  \def\calC{{\cal C}}

 \def\calR{{\cal R}}

\def\cNP{\hbox{\rm \sffamily NP}}

\def\cW#1{\hbox{\rm \sffamily W[#1]}}

\def\cFPT{\hbox{\rm \sffamily FPT}}

\def\eps{\varepsilon}
\def\O{\mathcal{O}{}}

\def\wlt{\vartriangleleft}

\def\inter#1{\left<#1\right>}

\def\lft{\leftarrow}
\def\rt{\rightarrow}
\def\lftrt{\leftrightarrow}
\def\nest{\nu}
\def\len{\lambda}
\def\pred{\textrm{\rm Pred}}

\def\recog{\textsc{Recog}\xspace}
\def\ext{\textsc{RepExt}\xspace}

\def\partition{\textsc{$3$-Partition}\xspace}
\def\binpacking{\textsc{BinPacking}\xspace}

\def\int{\hbox{\bf \rm \sffamily INT}\xspace}
\def\pint{\hbox{\bf \rm \sffamily PROPER INT}\xspace}
\def\uint{\hbox{\bf \rm \sffamily UNIT INT}\xspace}
\def\lint#1{\hbox{\bf \rm \sffamily $#1$-LengthINT}\xspace}
\def\nint#1{\hbox{\bf \rm \sffamily $#1$-NestedINT}\xspace}

\title{On the Classes of Interval Graphs of Limited Nesting\\and Count of Lengths\thanks{ The
conference version appeared in ISAAC 2016~\cite{kos}. The first author is supported by CE-ITI
(GA\v{C}R P202/12/G061) and Charles University as GAUK 1334217.}}

\author{Pavel Klav\'{\i}k \inst{1} \and
		Yota Otachi \inst{2} \and
		Ji\v{r}\'{\i} \v{S}ejnoha \inst{3}}

\institute{Computer Science Institute, Faculty of Mathematics and Physics, Charles University, Prague,\\
			Czech Republic. E-mail: \texttt{klavik@iuuk.mff.cuni.cz} \and
		School of Information Science, Japan Advanced Institute of Science and Technology. Japan.\\
			E-mail: \texttt{otachi@jaist.ac.jp} \and
		Department of Applied Mathematics, Faculty of Mathematics and Physics, Charles University, Prague,\\
			Czech Republic. E-mail: \texttt{jirka.sejnoha@gmail.com}}

\begin{document}

\maketitle

\begin{abstract}
In 1969, Roberts introduced \emph{proper} and \emph{unit interval graphs} and proved that these
classes are equal. Natural generalizations of unit interval graphs called \emph{$k$-length interval
graphs} were considered in which the number of different lengths of intervals is limited by $k$.
Even after decades of research, no insight into their structure is known and the complexity of
recognition is open even for $k=2$.  We propose generalizations of proper interval graphs called
\emph{$k$-nested interval graphs} in which there are no chains of $k+1$ intervals nested in each
other. It is easy to see that $k$-nested interval graphs are a superclass of $k$-length interval
graphs.

\hskip 2em We give a linear-time recognition algorithm for $k$-nested interval graphs. This
algorithm adds a missing piece to Gajarsk\'y et al.~[FOCS 2015] to show that testing FO properties
on interval graphs is \cFPT\ with respect to the nesting $k$ and the length of the formula, while
the problem is \cW2-hard when parameterized just by the length of the formula. We show that a
generalization of recognition called \emph{partial representation extension} is \cNP-hard for
$k$-length interval graphs, even when $k=2$, while Klav\'{\i}k et al. show that it is
polynomial-time solvable for $k$-nested interval graphs.
\medskip

\medskip

{\bfseries Diagram:} For a dynamic structural diagram of our results, see the following website
(supported Firefox and Google Chrome): \url{http://pavel.klavik.cz/orgpad/nest_len_int.html}
\end{abstract}

\tableofcontents

\section{Introduction}

For a graph $G$, we denote by $V(G)$ its vertices and $E(G)$ its edges.  An \emph{interval
representation} $\calR$ of a graph $G$ is a collection $\bigl\{\inter{u} : u \in V(G)\bigr\}$ of
intervals of the real line such that $uv \in E(G)$ if and only if $\inter{u} \cap \inter {v} \ne
\emptyset$. A graph is an \emph{interval graph} if it has an interval representation, and we denote
the class of interval graphs by \int.

An interval representation is called \emph{proper} if $\inter{u} \subseteq \inter{v}$ implies
$\inter{u} = \inter{v}$, and \emph{unit} if the length of all intervals $\inter{u}$ is one. The
classes of \emph{proper} and \emph{unit interval graphs} (denoted \pint\ and \uint) consist of all
interval graphs which have proper and unit interval representations, respectively.
Roberts~\cite{roberts_theorem} proved that $\pint = \uint$.

\begin{figure}[t!]
\centering
\includegraphics{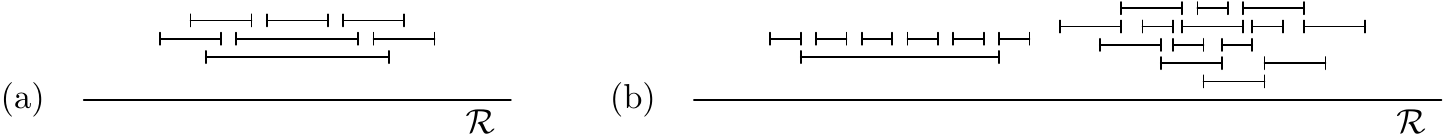}
\caption{(a) An interval representation with the nesting three. (b) The disjoint union of two
components with the minimum nesting two requiring three different lengths of intervals. On the
left, the shorter intervals are shorter than $1 \over 4$ of the longer ones. On the right, they are 
longer than $1 \over 3$.}
\label{fig:examples}
\end{figure}

\heading{The Studied Classes.} In this paper, we consider two hierarchies of subclasses of interval
graphs which generalize proper and unit interval graphs. The class \nint{k} consists of all interval
graphs which have representations with no $k+1$ intervals $\inter{u_0},\dots,\inter{u_k}$ such that
$\inter{u_0} \subsetneq \inter{u_1} \subsetneq \cdots \subsetneq \inter{u_k}$; see
Fig.~\ref{fig:examples}a.  The class \lint{k} consists of all interval graphs which have
representations having at most $k$ different lengths of intervals; see Fig.~\ref{fig:examples}b.  We
know by~\cite{roberts_theorem} that $\nint{1} = \pint = \uint = \lint{1}$.

\begin{figure}[b!]
\centering
\includegraphics{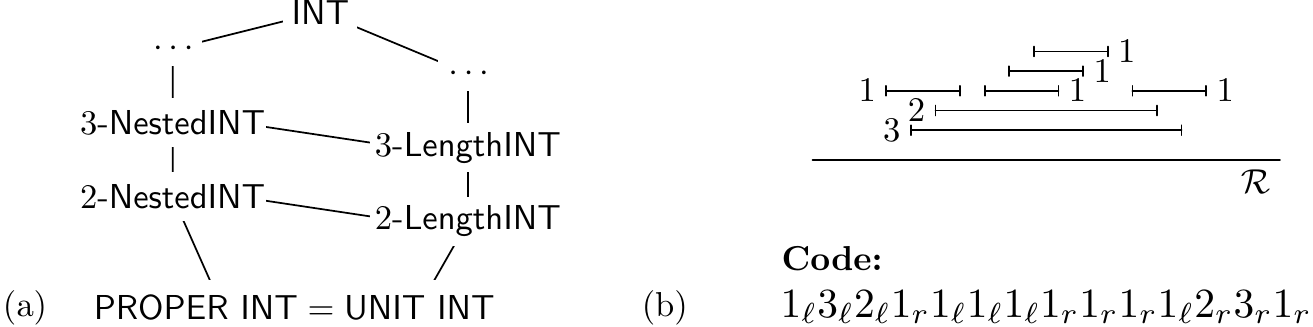}
\caption{(a) The Hasse diagram of proper inclusions of the considered classes. (b) We
can label each interval by the length of a maximal chain of nested intervals
ending in it. We code the graph by the left-to-right sequence of left endpoints $\ell$ and right
endpoints $r$ together with their labels.}
\label{fig:hierarchies}
\end{figure}

For an interval graph $G$, we denote the minimum nesting over all interval representations by
$\nest(G)$, and the minimum number of interval lengths by $\len(G)$.  Since nested intervals have different
lengths, we know that $\nest(G) \le \len(G)$ and this inequality may be strict (as in
Fig.~\ref{fig:examples}b). For each $k \ge 2$,
\begin{linenomath*}
$$\lint{(k-1)} \subsetneq \lint{k} \subsetneq \nint{k} \subsetneq \nint{(k+1)}.$$
\end{linenomath*}
Fishburn~\cite[Theorem 5, p. 177]{fishburn} shows that graphs $G$ in $\nint{2}$ have unbounded
$\len(G)$. Therefore, $\nint{2} \not\subseteq \lint{k}$ for each $k$. Figure~\ref{fig:hierarchies}a
depicts inclusions of considered classes.

\heading{Recognition.}
For a subclass $\calC$ of interval graphs, the following classical computational problem is studied:

\computationproblem
{Recognition -- $\recog(\calC)$}
{A graph $G$.}
{Is there a $\calC$-interval representation of $G$?}
{6.35cm}

\noindent The problem $\recog(\nint1) = \recog(\lint1)$ can be solved in linear time~\cite{uint_corneil}.

\heading{Partial Representation Extension.}
These problems generalizing recognition were introduced by Klav\'{\i}k et al.~\cite{kkv}.  A
\emph{partial representation} $\calR'$ of $G$ is an interval representation $\bigl\{\inter x' : x
\in V(G')\bigr\}$ of an induced subgraph $G'$ of $G$.  The vertices of $G'$ and the intervals of
$\calR'$ are called \emph{pre-drawn}.  A representation $\calR$ of $G$ \emph{extends} $\calR'$ if
and only if it assigns the same intervals to the vertices of $G'$: $\inter x = \inter x'$ for every
$x \in V(G')$. 

\computationproblem
{Partial Representation Extension -- $\ext(\calC)$}
{A graph $G$ and a partial representation $\calR'$ of an induced subgraph $G'$.}
{Is there a $\calC$-interval representation of $G$ extending $\calR'$?}
{10.95cm}

An $\O(n^2)$-time algorithm for $\ext(\int)$ was given in~\cite{kkv}. There are two different
linear-time algorithms~\cite{blas_rutter,kkosv} for this problem.  Minimal obstructions making
partial representations non-extendible are described in~\cite{ks}.  A linear-time algorithm for
proper interval graphs~\cite{kkorssv} and a quadratic-time algorithm for unit interval
graphs~\cite{soulignac} are known.
	
The partial representation extension problems were considered for other graph classes.
Polynomial-time algorithms are known for circle graphs~\cite{cfk}, and permutation and function
graphs~\cite{kkkw}.  The problems are \cNP-hard for chordal graphs~\cite{kkos} and contact
representations of planar graphs~\cite{contact_planar_ext}. The complexity is open for circular-arc
and trapezoid graphs.

\begin{figure}[t!]
\centering
\includegraphics{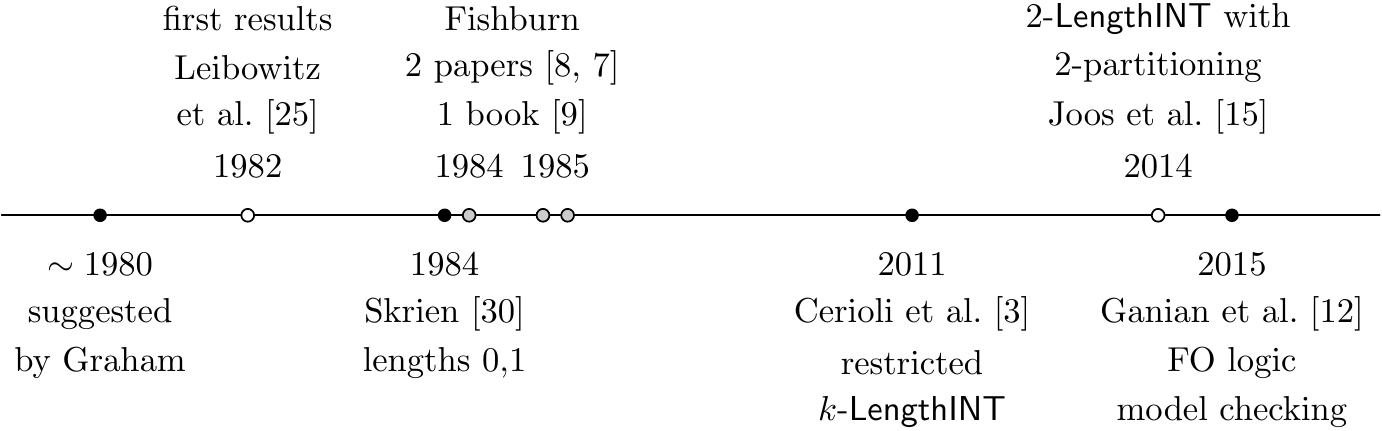}
\caption{Timeline of results for \lint{k}. Notice a big gap between 1985 and 2011.}
\label{fig:timeline}
\end{figure}

\heading{Previous Results and Motivation.}
The classes \lint{k} were introduced by Graham as a natural hierarchy between unit interval graphs
and interval graphs; see Fig.~\ref{fig:hierarchies}a. Unfortunately, even after decades, the only
results known are curiosities that illustrate the incredibly complex structure of \lint{k}, very
different from the case of unit interval graphs. For instance, \lint{k} is not closed under disjoint
unions; see Fig.~\ref{fig:examples}b. Timeline of results is depicted in Fig.~\ref{fig:timeline}.  

Leibowitz et al.~\cite{leibowitz_interval_count} show that the class \lint{2} contains caterpillars,
threshold graphs, and unit interval graphs with one additional vertex. Further, interval graphs $G$
with $\len(G) > 2$ such that $\len(G \setminus x) \le \len(G)-2$ for some $x \in V(G)$ are
constructed in~\cite{leibowitz_interval_count}.  Fishburn~\cite{fishburn_paper} shows that there are
infinitely many forbidden induced subgraphs for \lint{2}, while \lint{1} are interval graphs just
without $K_{1,3}$~\cite{roberts_theorem}. It is also known~\cite{fishburn_paradoxes} that there are
interval graphs in \lint{2} such that, when the shorter length is fixed to $1$, the longer one can
be one of the real numbers belonging to arbitrary many distinct intervals of the real line,
arbitrary far from each other.

Not much is known about the computational complexity of problems involving \lint{k}, even
recognition is open for $k=2$.  In~\cite{extended_bullfree}, a polynomial-time algorithm is given
for computing $\len(G)$ for interval graphs $G$ which are extended bull-free or almost threshold
(which highly restricts them).  Skrien~\cite{skrien} characterized \lint{2} which can be
realized by lengths zero (points) and one (unit intervals), leading to a linear-time recognition
algorithm.  As most of the efficient algorithms for intersection graph classes require
representations, very little is known how to algorithmically use that a given interval graph can be
represented by $k$ lengths.  In this paper, we show that partial representation extension is
\cNP-hard already for \lint{2}.

All these difficulties lead us to introduce another hierarchy of \nint{k} which generalizes proper
interval graphs; see Fig.~\ref{fig:hierarchies}.  We illustrate the nice structure of \nint{k} by
describing a relatively simple linear-time recognition algorithm based on MPQ-trees.  To the best of
our knowledge, the only reference is Fishburn's book~\cite{fishburn} in which the parameter
$\nest(G)$ called \emph{depth} is considered and linked to \lint{k}. There are some different
generalizations of proper interval graphs~\cite{proskurowski_telle}, which are less rich and not
linked to \lint{k}.

Since \nint{k} seem to share many properties with proper interval graphs, several future directions
of research are immediately offered. Using our results, it is possible to describe minimal forbidden
induced subgraphs~\cite{hkko}. For the computational problems which are tractable for proper
interval graphs and hard for interval graphs, the complexity of the intermediate problems for
\nint{k} can be studied.  (One such problem is FO property checking, discussed below.)
In Lemma~\ref{lem:nested_int_encoding}, we show that \nint{k} can be efficiently encoded, similarly
to proper interval graphs. 

\heading{Our Results.}
In~\cite{interval_count_two_partition}, a polynomial-time algorithm is given for recognizing
\lint{2} when intervals are partitioned into two subsets $A$ and $B$, each of one length, and both
$G[A]$ and $G[B]$ are connected. This approach might be generalized for partial representation extension,
but we show that removing the connectedness condition makes it hard:

\begin{theorem} \label{thm:repext_lint_hardness}
The problem $\ext(\lint{2})$ is \cNP-hard when every pre-drawn interval is of one length $a$.  It
remains \cNP-hard even when (i) the input prescribes two lengths $a=1$ and $b$, and (ii) for every
interval, the input assigns one of the lengths $a$ or $b$.  Also, it is \cW1-hard when parameterized
by the number of pre-drawn intervals.
\end{theorem}


We describe a dynamic programming algorithm for recognizing \nint{k}, based on a data structure called
an MPQ-tree. We show that we can optimize nesting greedily from the bottom to the top. We compute a
so-called minimal representation for each subtree and we show how to combine them.

\begin{theorem} \label{thm:recog_nint}
The minimum nesting number $\nest(G)$ can be computed in time $\O(n+m)$ where $n$ is the number of
vertices and $m$ is the number of edges. Therefore, the problem $\recog(\nint{k})$ can be solved
in linear time.
\end{theorem}

This result has the following application in the computational complexity of deciding logic formulas
over graphs. Let $\varphi$ be the length of a first-order logic formula for graphs. By the locality,
this formula can be decided in $G$ in time $n^{\O(\varphi)}$. Since it is \cW2-hard to decide it for
general graphs when parameterized by $\varphi$, it is natural to ask for which graph classes there
exists an FPT algorithm running in time $\O(n^c \cdot f(\varphi))$ for some computable function $f$.

In~\cite{fo_checking}, it is shown that the problem above is \cW2-hard even for interval graphs.  On
the other hand, if an interval graph is given together with a $k$-length interval
representation, \cite{fo_checking} gives an FPT algorithm with respect to the parameters $\varphi$
and the particular lengths of the intervals. It was not clear whether such an algorithm is
inherently geometrical.  Recently, Gajarsk\'y et al.~\cite{fo_posets} give a different FPT algorithm
for FO property testing for interval graphs parameterized by $\varphi$ and the nesting $k$, assuming
that a $k$-nested interval representation is given by the input. By our result, this assumption can
be removed since we can compute an interval representation of the optimal nesting in linear time.

The problem $\ext(\nint{k})$ is more involved since a straightforward greedy optimization from the
bottom to the top does not work. The described recognition algorithm can be generalized to solve
$\ext(\nint{k})$ in polynomial time~\cite{kos17b}.  It contrasts with
Theorem~\ref{thm:repext_lint_hardness}. The partial representation extension problems for \nint{k}
and \lint{k} are problems for which the geometrical version (at most $k$ lengths) is much harder
than the corresponding topological problem (the left-to-right ordering of endpoints of intervals).  

\section{Extending Partial Representations with Two Lengths}

The complexity of recognizing $\lint{k}$ is a long-standing open problem, even for $k=2$.  In this
section, we show that $\ext(\lint{k})$ is \cNP-hard even when $k=2$. We adapt the reduction from
\partition used in~\cite{kkos,kkorssv} which is the following computational problem:

\computationproblem
{\partition}
{Integers $A_1,\dots,A_{3s}$ and $M$ such that ${M \over 2} < A_i < {M \over 4}$ and $\sum A_i =
Ms$.} 
{Can $A_i$'s be split into $s$ triples, each summing to exactly $M$.}
{10.5cm}

\noindent This problem is strongly \cNP-complete~\cite{partition}, which means that it is \cNP-complete even
when the input is coded in unary, i.e., all integers are of polynomial sizes.

\begin{proof}[Theorem~\ref{thm:repext_lint_hardness}]
Assume (i) and (ii). For an instance of \partition, the reduction constructs an interval graph $G$
and a partial representation $\calR'$ as depicted in Fig.~\ref{fig:partition_reduction}.  We claim
that $\calR'$ can be extended using two lengths of intervals if and only if the instance of
\partition\ is solvable. We set $a=1$ and $b=s \cdot (M+2)-1$.  The partial representation $\calR'$
consists of $s+1$ disjoint pre-drawn intervals $\inter{v_0}',\dots,\inter{v_s}'$ of length $a$ such
that $\inter{v_i}' = [i \cdot (M+2), i \cdot (M+2)+1]$. So they split the real line into $s$ equal
gaps of size $M+1$ and two infinite regions.

\begin{figure}[b!]
\centering
\includegraphics{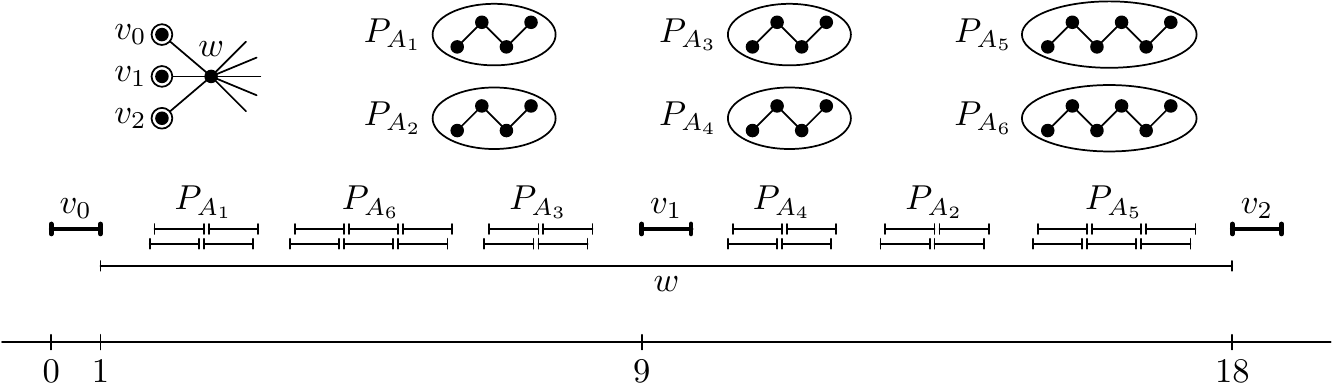}
\caption{Suppose that we have the following input for \partition: $s = 2$, $M = 7$, $A_1 = A_2 = A_3
= A_4 = 2$ and $A_5 = A_6 = 3$. The associated graph $G$ is depicted on top, and at the bottom we
find one of its extending representations, giving the 3-partitioning $\{A_1,A_3,A_6\}$ and
$\{A_2,A_4,A_5\}$.}
\label{fig:partition_reduction}
\end{figure}

Aside $v_0,\dots,v_s$, the graph $G$ contains a vertex $w$ represented by an interval of length $b$,
adjacent to every vertex in $G$.  Further, for each $A_i$, the graph $G \setminus w$ contains
$P_{2A_i}$ (a path with $2A_i$ vertices) as one component, with each vertex represented by an
interval of length $a$.

The described reduction clearly runs in polynomial time. It remains to show that $\calR'$ is
extendible if and only if the instance of \partition\ is solvable. The length of $b$ implies that
every extending representation has $\inter w = [1,s\cdot (M+2)]$ to intersect both $\inter{v_0}'$
and $\inter{v_s}'$. Therefore, each of the paths $P_{2A_i}$ has to be placed in exactly one of the
$s$ gaps. In every representation of $P_{2A_i}$, it requires the space at least $A_i+\eps$ for some
$\eps > 0$.  Three paths can be packed into the same gap if and only if their three integers sum to
at most $M$. Therefore, an extending representation $\calR'$ gives a solution to \partition, and
vice versa. A similar reduction from \binpacking implies $\cW1$-hardness when parameterized by the
number of pre-drawn intervals; see~\cite{kkos} for details.

This reduction can be easily modified when (i) and (ii) are avoided. We add two extra vertices:
$w_0$ adjacent to $v_0$ and $w_s$ adjacent to $v_s$, both non-adjacent to $w$. It forces the length
of $w$ to be in $[s \cdot (M+2)-1, s \cdot (M+2)+1)$, so the length $b$ does not have to be
prescribed. Also, this reduction works even when non-predrawn intervals do not have lengths
assigned.
\end{proof}

\section{Preliminaries and Basic Properties of k-Nested Interval Graphs} \label{sec:basic_properties}

In this section, we describe basic definitions and properties about nesting in interval
representations and about \nint{k}.

\heading{Definitions.}
For an interval representation $\calR$, the nesting defines a partial ordering $\subsetneq$ of
intervals. Intervals $\inter{u_1},\dots,\inter{u_k}$ form a chain of nested intervals of length $k$
if $\inter{u_1} \subsetneq \inter{u_2} \subsetneq \cdots \subsetneq \inter{u_k}$.  By $\nest(u)$, we
denote the length of a longest chain of nested intervals ending with $\inter u$. We denote
$\nest(\calR)$ the length of a longest chain of nested intervals in $\calR$, i.e.,
\begin{linenomath*}
$$
\nest(\calR) = \max_{u \in V(G)} \nest(u)\qquad\text{and}\qquad
\nest(G) = \min_\calR \nest(\calR) = \min_\calR \max_{u \in V(G)} \nest(u).
$$
\end{linenomath*}
For $A \subseteq V(G)$, we denote by $G[A]$ the subgraph of $G$ induced by $A$. For a representation
$\calR$ of $G$, let $\calR[A]$ be the representation of $G[A]$ created by restricting $\calR$ to
the intervals of $A$. And for an induced subgraph $H$ of $G$, let $\calR[H] = \calR[V(H)]$.

\heading{Pruning Twins.}
Two vertices $x$ and $y$ are twins if and only if $N[x] = N[y]$. The standard observation is that
twins can be ignored since they can be represented by identical intervals, and notice that this does
not increase nesting and the number of lengths. We can locate all twins in time
$\O(n+m)$~\cite{recog_chordal_graphs} and we can \emph{prune} the graph by keeping one vertex per
equivalence class of twins. An interval graph belongs to \nint{k} if and only if the pruned graph
belongs to \nint{k}.

\heading{Decomposition into Proper Interval Representations.}
The following equivalent definition of \nint{k} is used by Gajarsk\'y et al.~\cite{fo_posets}:

\begin{lemma} \label{lem:k_nested_partitioning}
An interval graph belongs to \nint{k} if and only if it has an interval representation which can be
partitioned into $k$ proper interval representations.
\end{lemma}

\begin{proof}
Let $\calR$ be an interval representation partitioned into proper interval representations
$\calR_1,\dots,\calR_k$. No chain of nested intervals contains two intervals from some $\calR_i$, so
the nesting is at most $k$. On the other hand, let $\calR$ be a $k$-nested interval representation.
We label each interval $\inter u$ by $\nest(u)$; see Fig.~\ref{fig:hierarchies}b.  Notice that the
intervals of each label $i \in \{1,\dots,k\}$ form a proper interval representation $\calR_i$.
\end{proof}

\heading{Efficient encoding.}
An interval graph can be encoded by $2n \lceil\log n\rceil$ bits by labeling the vertices
$1,\dots,n$ and listing the left-to-right ordering of labels of the endpoints. Proper interval
graphs can be encoded more efficiently using only $2n$ bits: the sequence of endpoints ($\ell$ for
left one, $r$ for right one), as they appear from left to right. We generalize it for
$\nint{k}$.

\begin{lemma} \label{lem:nested_int_encoding}
A graph in \nint{k} can be encoded by $2n \lceil\log k + 1\rceil$ bits where $n$ is the number of
vertices.
\end{lemma}

\begin{proof}
See Fig.~\ref{fig:hierarchies}b for an example. Let $\calR_1,\dots,\calR_k$ be the labeling from the
proof of Lemma~\ref{lem:k_nested_partitioning}. From left to right, we output $\ell$ or $r$ for each
endpoint together with its labels. This encoding takes $\lceil \log k + 1 \rceil$ bits per endpoint.
\end{proof}

\heading{Minimal Forbidden Induced Subgraphs.}
Interval graphs and the subclasses \nint{k} and \lint{k} are closed under induced
subgraphs, so they can be characterized by minimal forbidden induced subgraphs. Lekkerkerker and
Boland~\cite{lb_graphs} describe them for interval graphs, and Roberts~\cite{roberts_theorem} proved
that $\nint{1}=\lint{1}$ are claw-free interval graphs. On the other hand, \lint{2} have
infinitely many minimal forbidden induced subgraphs~\cite{fishburn_paper} which are interval graphs.
In~\cite{hkko}, our results in Section~\ref{sec:recognizing_nesting} are used to describe minimal
forbidden induced subgraphs for \nint{k}.

\section{Maximal Cliques and MPQ-trees} \label{sec:maximal_cliques}

In this section, we review well-known properties of interval graphs. First, we describe their
characterization in terms of orderings of maximal cliques. Then we introduce a data structure called
an MPQ-tree which stores all feasible orderings.

\begin{figure}[b!]
\centering
\includegraphics{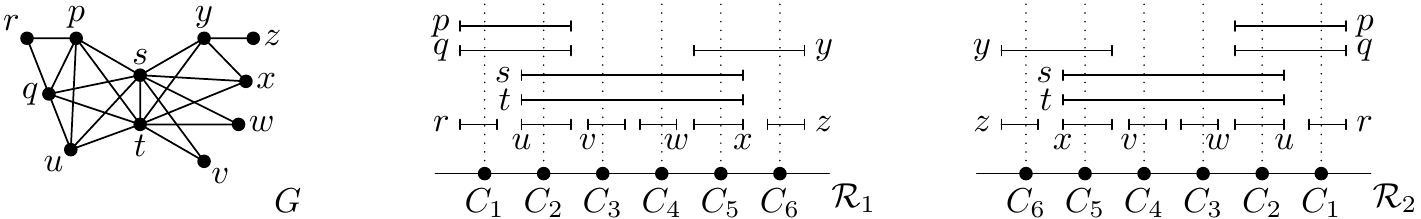}
\caption{An interval graph $G$ and two of its representations with different left-to-right orderings of
the maximal cliques, with choices of clique-points.}
\label{fig:fulkerson_gross}
\end{figure}

\heading{Consecutive Orderings.}
Fulkerson and Gross~\cite{maximal_cliques} proved the following characterization of interval
graphs; see Fig.~\ref{fig:fulkerson_gross}:

\begin{lemma}[Fulkerson and Gross~\cite{maximal_cliques}] \label{lem:maximal_cliques}
A graph is an interval graph if and only if there exists a linear ordering $<$ of its maximal
cliques such that, for each vertex, the maximal cliques containing this vertex appear consecutively.
\end{lemma}

An ordering of the maximal cliques satisfying the statement of Lemma~\ref{lem:maximal_cliques} is
called a \emph{consecutive ordering}. Each interval graph has $\O(n)$ maximal cliques of total size
$\O(n+m)$ which can be found in linear time~\cite{recog_chordal_graphs}.

\begin{figure}[b!]
\centering
\includegraphics{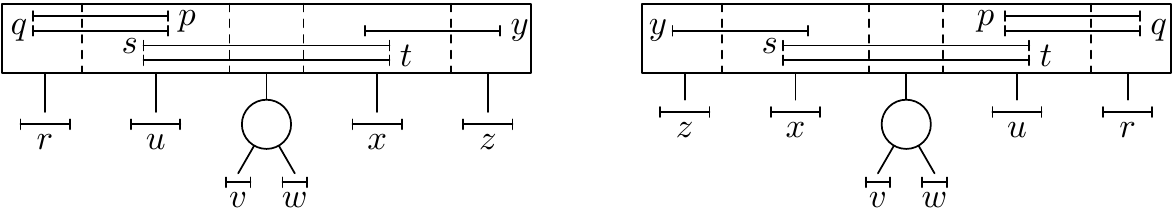}
\caption{Two equivalent MPQ-trees with denoted sections. In all figures, we denote P-nodes by
circles and Q-nodes by rectangles.}
\label{fig:pq_trees}
\end{figure}

\heading{Cleaned Representations.}
For a given consecutive ordering of maximal cliques, it is easy to construct a representation
the number of all nestings called a \emph{cleaned representation}. 

\begin{lemma} \label{lem:cleaned_representations}
For a given consecutive ordering $<$ of maximal cliques, there exists a cleaned representation
such that if $\inter u \subsetneq \inter v$, then $\inter u$ is nested in $\inter v$ in every 
interval representation with this consecutive ordering $<$. We can construct it in time $\O(n)$.
\end{lemma}

\begin{proof}
We place maximal cliques on the real line according to $<$. For each $v \in V(G)$, we place $\inter
v$ on top of the maximal cliques containing $v$.  Let $v^\lft$ be the left-most clique containing
$v$ and $v^\rt$ be the right-most clique containing $v$. We place $\inter v$ on the left of $v^\lft$
and on the right of $v^\rt$.

For a maximal clique $C$, let $u_1,\dots,u_\ell$ be all vertices having $u_i^\lft = C$, i.e., all
intervals $\inter{u_i}$ start at $C$. Since there are no twins, we have $u_i^\rt \ne u_j^\rt$ for
all $i \ne j$. We order the left endpoints of $\inter{u_1},\dots,\inter{u_\ell}$ exactly as the
maximal cliques $u_1^\rt,\dots,u_\ell^\rt$ are ordered in $<$.  Similarly, let $v_1,\dots,v_{\ell'}$
be all vertices having $v_i^\rt = C$. We order the right endpoins of
$\inter{v_1},\dots,\inter{v_{\ell'}}$ exactly as the maximal cliques $v_1^\lft,\dots,v_{\ell'}^\lft$
are ordered in $<$.

The constructed interval representation avoids all unnecessary nesting. We get that $\inter u
\subsetneq \inter v$ if and only if $v^\lft < u^\lft \le u^\rt < v^\rt$ in which case the nesting is
clearly forced by the consecutive ordering $<$.  The construction clearly runs in time $\O(n+m)$.
\end{proof}

\heading{PQ-trees.} A \emph{PQ-tree} $T$ is a rooted tree, introduced by Booth and
Lueker~\cite{pq_trees}. Its leaves are in one-to-one correspondence with the maximal cliques. Its
inner nodes are of two types: \emph{P-nodes} and \emph{Q-nodes}. Each P-node has at least two
children, each Q-node at least three.  The orderings of the children of inner nodes are given. The
PQ-tree $T$ represents one consecutive ordering $<_T$ called the \emph{frontier} of $T$ which is the
ordering of the leaves from left to right.  

The PQ-tree $T$ represents all consecutive orderings of $G$ as frontiers of equivalent PQ-trees
which can be constructed from $T$ by sequences of \emph{equivalent transformations} of two types:
(i) an arbitrary reordering of the children of a P-node, and (ii) a reversal of the order of the
children of a Q-node; see Fig.~\ref{fig:pq_trees}. 

A \emph{subtree} $T'$ of the PQ-tree $T$ consists of a node and all its descendants.  For a node
$N$, we denote the subtree having $N$ as the root by $T[N]$ and its subtrees are the subtrees which
have the children of $N$ as the roots.

\heading{MPQ-trees.} An \emph{MPQ-tree}~\cite{korte_mohring} is an augmentation of a PQ-tree $T$ in
which the nodes of $T$ have assigned subsets of $V(G)$ called \emph{sections}.  To a leaf
representing a clique $C$, we assign one section $s(C)$.  Similarly, to each P-node $P$, we assign
one section $s(P)$. For a Q-node $Q$ with subtrees $T_1,\dots,T_q$, we have $q$ sections
$s_1(Q),\dots,s_q(Q)$ ordered from left to right, each corresponding to one subtree, and let $s(Q) =
s_1(Q) \cup \dots \cup s_q(Q)$.  Examples of sections are depicted in Fig.~\ref{fig:pq_trees}.

The section $s(C)$ has all vertices contained in the maximal clique $C$ and no other maximal clique.
The section $s(P)$ of a P-node $P$ has all vertices that are contained in all maximal cliques of the
subtree rooted at $P$ and in no other maximal clique.  Let $Q$ be a Q-node with subtrees
$T_1,\dots,T_q$. Let $x$ be a vertex contained only in maximal cliques of the subtree rooted at $Q$,
contained in maximal cliques of at least two subtrees. Then $x$ is contained in every section
$s_i(Q)$ such that some maximal clique of $T_i$ contains $x$.

Every vertex $x$ is in sections of exactly one node of $T$. In the case of a Q-node, it is placed in
consecutive sections of this node.  For a Q-node $Q$, if $x$ is placed in a section $s_i(Q)$, then
$x$ is contained in all cliques of $T_i$. Every section of a Q-node is non-empty, and two
consecutive sections have a non-empty intersection.  After pruning twins, no two vertices belong to
the exactly same sections of the MPQ-tree.

Let $G[T]$ be the interval graph induced by the vertices of the sections of $T$.  By $G[N]$, we
denote $G[T[N]]$. For a representation $\calR$, we have $\calR[T] = \calR[G[T]]$ and $\calR[N] =
\calR[T[N]]$.

\heading{Forced Nestings.}
Let $Q$ be a Q-node with sections $s_1(Q),\dots,s_q(Q)$ and $u$ be a
vertex.
\begin{packed_itemize}
\item If $u$ does not belong to sections of $T[Q]$, let $s_u^\lft(Q) = s_1(Q)$ and $s_u^\rt(Q) =
s_q(Q)$.
\item If $u \in s(Q)$, let $s_u^\lft(Q)$ be the leftmost section of $Q$ containing $u$ and
$s_u^\rt(Q)$ be the rightmost one.
\item If $u$ belongs to sections of a subtree $T_i$ of $Q$, we put $s_u^\lft(Q) = s_u^\rt(Q) = s_i(Q)$. 
\end{packed_itemize}

We study under which conditions is $\inter u$ forced to be nested in $\inter v$ in every interval
representation, and we represent this by a partial ordering $\subsetneq_F$.  We have $u \subsetneq_F
v$ if and only if there there exists a Q-node $Q$ such that $s_v^\lft(Q)$ is on the left of
$s_u^\lft(Q)$ and $s_v^\rt(Q)$ is on the right of $s_u^\rt(Q)$.

\begin{lemma} \label{lem:forced_nestings}
We have $\inter u \subsetneq \inter v$ for every interval representation if and only if $u
\subsetneq_F v$.
\end{lemma}

\begin{proof}
If $u \subsetneq_F v$, then every consecutive ordering contains at least one maximal clique
containing $v$ on the left of all maximal cliques containing $u$ and at least one on the right, so
necessarily $\inter u \subsetneq \inter v$.

Suppose that there exists a cleaned representation with $\inter u \subsetneq \inter v$. Therefore, every
maximal clique contaning $u$ also contains $v$, so $u$ and $v$ appear in sections of a path from
a leaf to the root of the MPQ-tree, and $v$ appears at least as high as $u$. Suppose that $u
\not\subsetneq_F v$. Both $u$ and $v$ do not belong to a same Q-node, otherwise they could not be
nested in a cleaned representation. There is no Q-node on the
path between $u$ and $v$, above $u$; possibly $u$ belongs to all sections of one Q-node. Therefore,
we can reorder all these P-nodes to place the subtrees containing $u$ on the side, and the obtained
cleaned representation has $\inter u \not\subsetneq \inter v$.
\end{proof}

\section{Recognizing k-nested Interval Graphs} \label{sec:recognizing_nesting}

In this section, we describe a linear-time algorithm for computing minimal nesting of interval
graphs. By Lemma~\ref{lem:cleaned_representations}, the problem reduces to finding a consecutive
ordering of maximal cliques which minimizes the nesting of a cleaned representation. So we want to
reorder the MPQ-tree to minimize the nesting, which is done by dynamic programming from the bottom
to the top.

\begin{figure}[t!]
\centering
\includegraphics{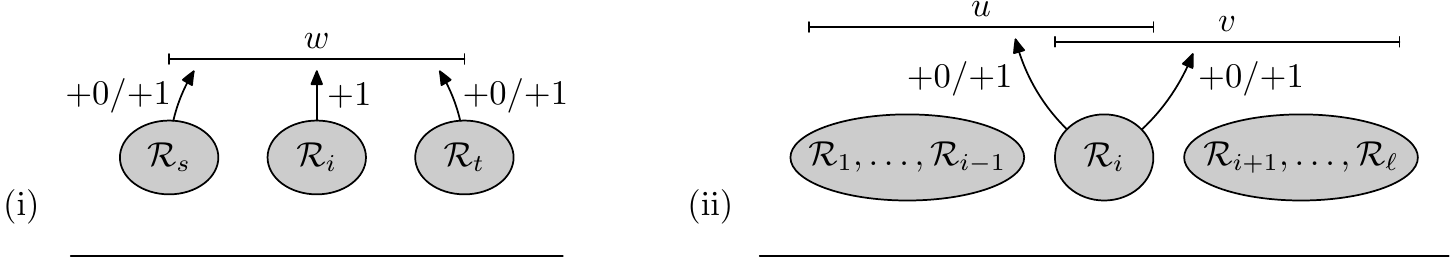}
\caption{(i) The nesting $\nest(G[T_i])$, for $i \ne s,t$, is always increased by one with $\inter
w$, but the nestings $\nest(G[T_s])$ and $\nest(G[T_t])$ may or may not be increased by one. (ii)
The nesting $\nest(G[T_i])$ may be increased by one with $\inter u$ or $\inter v$. It might not be
possible to preserve nesting on both sides, for instance when $G[T_i]$ is the disjoint union of
$K_{1,3}$ and $K_1$.}
\label{fig:intuition}
\end{figure}

\heading{Intuition.}
We process the MPQ-tree from the bottom to the top, and we optimize the nesting. Let $N$ be a node
of the MPQ-tree and let $T_1,\dots,T_\ell$ be its subtrees. Suppose that we know
$\nest(G[T_1]),\dots,\nest(G[T_\ell])$ from the dynamic programming.  Is $\nest(G[N])$ determined?
The answer is that almost. Let $\calR_1,\dots,\calR_\ell$ be interval representations of
$G[T_1],\dots,G[T_\ell]$ minimizing the nesting. We consider two model situations, depicted in
Fig.~\ref{fig:intuition}:
\begin{packed_enum}
\item[(i)] Suppose that $N$ is a P-node with $s(N) = \{w\}$. Then $G[N]$ is the disjoint union of
$G[T_1],\dots,G[T_\ell]$ together with the universal vertex $w$. Every interval representation of
$G[N]$ looks as depicted in Fig.~\ref{fig:intuition}i. We have two representations $\calR_s$ and
$\calR_t$ placed on the left and right sides of $\inter w$, respectively, while the remaining
$\calR_i$, for $i \ne s,t$, are placed inside $\inter w$.  Therefore, their nestings $\nest(G[T_i])$
are increased by one with $\inter w$. On the other hand, some intervals of $\calR_s$ and $\calR_t$
may stretch out of $\inter w$, so the nestings $\nest(G[T_s])$ and $\nest(G[T_t])$ is not
necessarily increased by one.  More precisely, the intervals of $\calR_t$ contained in the left-most
clique and the intervals of $\calR_s$ are not nested in $\inter w$ in a cleaned representation.
\item[(ii)] Suppose that $N$ is a Q-node and we consider the following simplified situation depicted
in Fig.~\ref{fig:intuition}ii. The graph $G[N]$ consists of $G[T_i]$ together with two universal
vertices $u$ and $v$, each attached some other part of $G[N]$ non-adjacent to all vertices of
$G[T_i]$. Then $\calR_i$ is covered from, say, left by $\inter u$ and from right by $\inter v$.  The
nesting of $\nest(G[T_i])$ is not necessarily increased by one with $\inter u$ or $\inter v$. More
precisely, in a cleaned representation, the intervals of $\calR_i$ contained in the left-most clique
are not nested in $\inter v$ and those contained in the right-most clique are not nested in $\inter
u$. It is possible that both sides cannot be optimized simultaneously. 
\end{packed_enum}
Therefore, the dynamic programming computes three values for each subtree $T$, denoted as a triple
$(\alpha,\beta,\gamma)$, which we define formally in the next subsection. We have $\alpha =
\nest(G[T])$. The value $\beta$ is the increase in the nesting when $T$ is placed on the side, as in
(i); so either $\beta = \alpha$, or $\beta = \alpha-1$. The value $\gamma$ is the increase in the
nesting of one side, subject to the other side being optimized according to $\beta$, as in (ii). So
always $\beta \le \gamma$ and either $\gamma = \alpha$ or $\gamma = \alpha-1$.

\subsection{Triples $(\alpha,\beta,\gamma)$}

\begin{figure}[b!]
\centering
\includegraphics{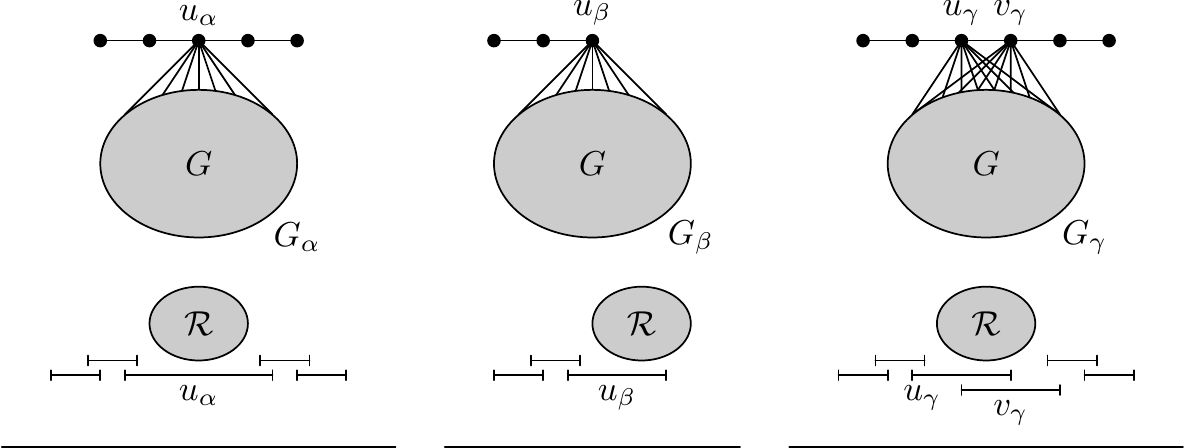}
\caption{The graphs $G_\alpha$, $G_\beta$ and $G_\gamma$ with representations, defining the triple
$(\alpha,\beta,\gamma)$ of $T$. The vertices of $G$ are adjacent to the added vertices $u_\alpha$,
$u_\beta$, $u_\gamma$, and $v_\gamma$, and not to the others. In bottom, we depict the structure of
their representations with $\calR$ being a representation of $G$.}
\label{fig:triple_graphs}
\end{figure}

For an interval graph $G$, we define the triple $(\alpha,\beta,\gamma)$ as follows. Let $G_\alpha$,
$G_\beta$ and $G_\gamma$ be the graphs constructed from $G$ as in Fig.~\ref{fig:triple_graphs}. Let
\begin{linenomath*}
$$
\alpha = \nest(G_\alpha)-1,\qquad
\beta = \nest(G_\beta)-1,\qquad\text{and}\qquad
\gamma = \nest(G_\gamma)-1.
$$
\end{linenomath*}
Similarly, for a subtree $T$ of the MPQ-tree, we define its triple as the triple of $G[T]$. 
The dynamic algorithm computes triples of all subtrees from the leaves to the root, and outputs $a$
of the root as $\nest(G)$. 

\begin{lemma} \label{lem:triple_bounds}
For every interval graph $G$, its triple $(\alpha,\beta,\gamma)$ satisfies $\alpha-1 \le \beta \le
\gamma \le \alpha$.
\end{lemma}

\begin{proof}
We prove equivalently that $\nest(G_\alpha)-1 \le \nest(G_\beta) \le \nest(G_\gamma) \le
\nest(G_\alpha)$. We trivially know that $\nest(G_\beta) \le \nest(G_\gamma)$ since $G_\beta$
is an induced subgraph of $G_\gamma$.

The definition of $G_\alpha$ implies that $\nest(G_\alpha) = \nest(G)+1$, since in every interval
representation of $G_\alpha$, both endpoints of $\inter{u_\alpha}$ are covered by attached paths,
and therefore a representation $\calR$ of $G$ is nested in $\inter{u_\alpha}$. Since $G$ is an
induced subgraph of $G_\beta$, we have $\nest(G) \le \nest(G_\beta)$, so the inequality
$\nest(G_\alpha) - 1 \le \nest(G_\beta)$ follows. For an alternative proof, consider a
representation of $G_\beta$ minimizing nesting. We modify it to a representation of $G_\alpha$ by
stretching $\inter{u_\beta}$ into $\inter{u_\alpha}$, which increases nesting by at most one, and by
adding the second path attached to $\inter{u_\alpha}$. So $\nest(G_\alpha) \le \nest(G_\beta) + 1$.

It remains to show the last inequality that $\nest(G_\gamma) \le \nest(G_\alpha)$. Consider a
representation of $G_\alpha$ with minimal nesting, we have $G$ strictly contained inside
$\inter{u_\alpha}$. By shifting $r(u_\alpha)$ to the left, we get $\inter{u_\gamma}$. By adding
$\inter{v_\gamma}$, we do not increase the nesting and we get a representation of $G_\gamma$. So
$\nest(G_\gamma) \le \nest(G_\alpha)$.
\end{proof}

\begin{figure}[t!]
\centering
\includegraphics{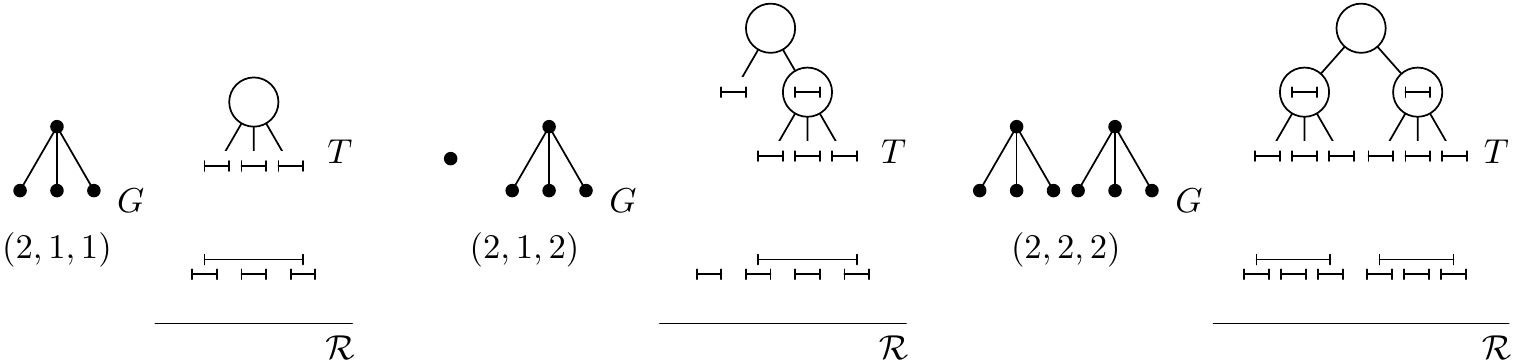}
\caption{Three interval graphs $G$ with $\nest(G) = 2$, together with MPQ-trees $T$ and
representations $\calR$ minimizing the nesting.}
\label{fig:three_types_of_triples}
\end{figure}

Therefore, the triples classify interval graphs into three types; see
Fig.~\ref{fig:three_types_of_triples} for examples.

\begin{corollary}
Interval graphs $G$ with $\nest(G) = k$ have triples of three types: $(k,k-1,k-1)$, $(k,k-1,k)$ and
$(k,k,k)$.
\end{corollary}

\heading{Interpreting Triples.}
Let $(\alpha,\beta,\gamma)$ be the triple for $G$. We want to argue how the formal
definition relates to the description in the last paragraph of Intuition. We can interpret the
triple of $G$ as increase in the nesting, depending how $G$ is represented with respect to the
rest of the graph. Since $\alpha = \nest(G)$, it is easy to understand. Next, we describe an
interpretation for the value $\beta$.

\begin{lemma} \label{lem:meaning_of_beta}
For every representation of $G_\beta$, we have $\nest(u_\beta) \ge \beta+1$.
\end{lemma}

\begin{proof}
We assume that a representation $\calR_\beta$ of $G_\beta$ is cleaned; it only decreases nesting.  By the
definition of $\beta$, there exists a maximal chain of nested intervals of length at least $\beta+1$.
Suppose that its length is at least two. Let $\calR = \calR_\beta[G]$, and we assume that
$\inter{u_\beta}$ covers $\calR$ from the left.  If this chain does not end with $\inter{u_\beta}$, it ends
with an interval of $\calR$ placed in the right-most maximal clique. Since every other interval of
the chain is nested in $\inter{u_\beta}$, we replace this end with $\inter{u_\beta}$, and obtain a chain of
nested intervals of length at least $\beta+1$ ending with $\inter{u_\beta}$.
\end{proof}

In other words, in every representation of $G$, there exists a chain of length at least $\beta$ which
is nested in any interval in the rest of the graph which plays the role of $\inter{u_\beta}$. In
Lemma~\ref{lem:minimal_representation}, we show that there exists a representation for which the
length of a longest such chain is exactly $\beta$. This links the value $\beta$ to
Fig.~\ref{fig:intuition}i.

Last, we describe an interpretation for the value $\gamma$.

\begin{lemma} \label{lem:meaning_of_gamma}
For every representation of $G_\gamma$, we have
\begin{linenomath*}
$$
\min \bigl\{\nest(u_\gamma),\nest(v_\gamma)\bigr\} \ge b+1
\qquad \text{and} \qquad
\max \bigl\{\nest(u_\gamma),\nest(v_\gamma)\bigr\} \ge c+1.
$$
\end{linenomath*}
\end{lemma}

\begin{proof}
We prove this similar as in Lemma~\ref{lem:meaning_of_beta}. Consider a cleaned representation
$\calR_\gamma$ of $G_\gamma$. It contains a maximal chain of length at least $\gamma+1$ ending with
$\inter x$.  If $x \ne u_\gamma$ and $x \ne v_\gamma$, we can replace $\inter x$ with both
$\inter{u_\gamma}$ and $\inter{v_\gamma}$, so both $\nest(u_\gamma) \ge \gamma+1$ and
$\nest(v_\gamma) \ge \gamma+1$. Otherwise, suppose that, say, $x = v_\gamma$. Then $\nest(v_\gamma)
\ge \gamma+1$ and by removing $\inter{v_\gamma}$ and the added intervals, we obtain a representation
of $\calR_\beta$ with $u_\beta = u_\gamma$. By Lemma~\ref{lem:meaning_of_beta}, $\nest(u_\beta) \ge
\beta+1$.
\end{proof}

Therefore, in every representation of $G$, there exists a chain of length at least $\gamma$ which is
nested in any interval in the rest of the graph which plays the role of either $\inter{u_\gamma}$ or
$\inter{v_\gamma}$, while there is a chain of length at least $\beta$ which in nested in any interval playing
the role of the other one. In Lemma~\ref{lem:minimal_representation}, we show that there exists a
representation for which the lengths of longest such chains are exactly $\beta$ and $\gamma$, respectively.
This links the value $\gamma$ to Fig.~\ref{fig:intuition}ii.

\heading{Minimal Representations.}
Let $(\alpha,\beta,\gamma)$ be a triple of an interval graph $G$ and let $\calR$ be a cleaned
representation of $G$ with $C^\lft$ and $C^\rt$ being the leftmost and the rightmost maximal cliques
in its consecutive ordering of maximal cliques. We define:
\begin{linenomath*}
$$
\nest^\rt(\calR) = \!\!\!\!\!\max_{x \in V(G) \setminus C^\lft}\!\!\!\!\! \nest(x),\qquad \text{and}\qquad
\nest^\lft(\calR) = \!\!\!\!\!\max_{x \in V(G) \setminus C^\rt}\!\!\!\!\! \nest(x).
$$
\end{linenomath*}
The representation $\calR$ of $G$ is \emph{minimal} if $\nest(\calR) = \alpha$, $\nest^\rt(\calR) =
\beta$ and $\nest^\lft(\calR) = \gamma$. So a minimal representation $\calR$ can be used
simultaneously in representations of $G_\alpha$, $G_\beta$ and $G_\gamma$ to get nestings
$\alpha+1$, $\beta+1$ and $\gamma+1$, respectively.  For instance, all representations in
Fig.~\ref{fig:three_types_of_triples} are minimal. 

\begin{lemma} \label{lem:minimal_representation}
For every interval graph $G$, there exists a minimal representation $\calR$.
\end{lemma}

\begin{proof}
We argue according to the type of the triple of $G$.

\emph{The triple $(k,k-1,k-1)$.} Let $\calR_\gamma$ be a cleaned representation of $G_\gamma$ minimizing the
nesting, we have $\nest(\calR_\gamma) = k$. Since $\nest(G) = k$, we have $\nest(\calR_\gamma[G]) = k$ as
well. By Lemma~\ref{lem:meaning_of_gamma}, $\nest(u_\gamma) \ge k$ and $\nest(v_\gamma) \ge k$, so we get
equalities. The representation $\calR = \calR_\gamma[G]$ has $\nest^\lft(\calR) = \nest^\rt(\calR)
= k-1$ and $\calR$ is minimal.

\emph{The triple $(k,k-1,k)$.} Let $\calR_\beta$ be a cleaned representation of $G_\beta$ minimizing
the nesting such that $\inter{u_\beta}$ intersects $\calR_\beta[G]$ from left, we have
$\nest(\calR_\beta) = k$. Let $\calR = \calR_\beta[G]$. Since $\nest(G) = k$, we have $\nest(\calR)
= k$ as well. Similarly as in the proof of Lemma~\ref{lem:meaning_of_beta}, we get $\nest^\rt(\calR)
= k-1$.  If $u_\gamma = u_\beta$ and we add $\inter{v_\gamma}$ with the attached path, we obtain a
representation of $G_\gamma$ with nesting at least $k+1$.  Therefore, $\nest^\lft(\calR) = k$ and
$\calR$ is minimal.

\emph{The triple $(k,k,k)$.} Let $\calR$ be a cleaned representation of $G$ minimizing the nesting,
so $\nest(\calR) = k$. If $\nest^\rt(\calR) < k$ or $\nest^\lft(\calR) < k$, we can add
$\inter{u_b}$ and the attached intervals to obtain a representation of $G_b$ of nesting $k$, so $b =
k-1$; a contradiction. So $\nest^\rt(\calR) = \nest^\lft(\calR) = k$ and $\calR$ is minimal.
\end{proof}

For a representation $\calR$, the \emph{flipped} representation $\calR^\lftrt$ is created by reversing the
left-to-right order of endpoints of intervals. Notice that $\nest(\calR) = \nest(\calR^\lftrt)$,
$\nest^\rt(\calR) = \nest^\lft(\calR^\lftrt)$ and $\nest^\lft(\calR) = \nest^\rt(\calR^\lftrt)$.

\begin{lemma} \label{lem:local_minimality}
For every interval graph $G$, there exists a cleaned representation $\calR$ of $G$ minimizing the
nesting such that for every subtree $T$ of its MPQ-tree, $\calR[T]$ is minimal or
$\calR^\lftrt[T]$ is minimal.
\end{lemma}

\begin{proof}
Let $\calR$ be a cleaned representation of $G$ minimizing the nesting, and consider a maximal
subtree $T$ for which $\calR[T]$ is not minimal. By Lemma~\ref{lem:minimal_representation}, there
exists a minimal representation $\calR_T$ of $G[T]$.  If $\nest^\rt(\calR[T]) \le
\nest^\lft(\calR[T])$, let $\calR_T^* = \calR_T$, otherwise let $\calR_T^* = \calR_T^\lftrt$. Since
$\calR[T]$ appears consecutively in $\calR$, we replace it by $\calR_T^*$, and construct a modified
cleaned representation $\hat\calR$ of $G$. It remains to argue that for every subtree $T'$
containing $T$, the representation $\hat\calR[T']$ remains minimal; the lemmas then follows by
induction.

We know that $\calR[T']$ is minimal. The modification only changed chains which start in
$\calR_T^*$. By Lemmas~\ref{lem:meaning_of_gamma}, \ref{lem:minimal_representation}, we get that
$\nest(\calR_T^*) \le \nest(\calR[T])$, $\nest^\rt(\calR_T^*) \le \nest^\rt(\calR[T])$ and
$\nest^\lft(\calR_T^*) \le \nest^\lft(\calR[T])$.  Therefore, every chain above $\calR[T]$ extends only 
chains with lengths equal or shorter, so $\hat\calR[T']$ remains minimal.
\end{proof}

\heading{Triples for Leaves.}
Recall that we have no twins. For a leaf $L$ of the MPQ-tree, we have either $G[L]$ having no
vertices (when $s(L) = \emptyset$), or $G[L] \cong K_1$ (when $s(L) = \{w\}$). In the former case,
the triple of $L$ is equal $(0,0,0)$. In the latter case, it is equal $(1,0,0)$.

\subsection{Triples for P-nodes}

\begin{figure}[t!]
\centering
\includegraphics{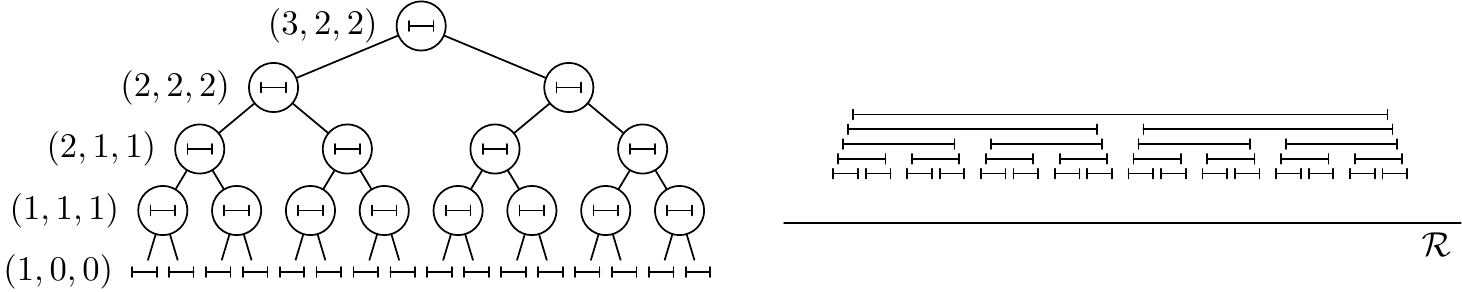}
\caption{An MPQ-tree representing $G$ with the computed triples $(\alpha,\beta,\gamma)$ (equal on
each level) and a cleaned representation minimizing nesting. We have that $\nest(G) = 3$.}
\label{fig:p_node_example}
\end{figure}

Let $T_1,\dots,T_p$ be the children of a P-node $P$, with $p \ge 2$, with the computed triple
$(\alpha_i,\beta_i,\gamma_i)$ for each subtree $T_i$. We compute the triple $(\alpha,\beta,\gamma)$
of the subtree $T = T[P]$ using the following formulas; see Fig.~\ref{fig:p_node_example} for an
example:
\begin{linenomath*}
\begin{eqnarray*}
\alpha &=&
\begin{cases}
\max \{\alpha_1,\dots,\alpha_p\},&\qquad \text{if } s(P) = \emptyset,\\
\min_{s \ne t} \max \{\beta_s+1,\beta_t+1,\alpha_i+1 : i \ne s,t\},&\qquad \text{if } s(P) = \{w\}.\\
\end{cases}\\
\beta &=& \min_{s} \max \{\beta_s, \alpha_i : i \ne s\},\\
\gamma &=& \max \{\alpha_1,\dots,\alpha_p\}.
\end{eqnarray*}
\end{linenomath*}
\vskip -1ex

\begin{lemma} \label{lem:p_node_formulas}
The formulas compute the triple $(\alpha,\beta,\gamma)$ of $T[P]$ correctly.
\end{lemma}

\begin{proof}
This proof also explains how these formulas are formed.

\emph{The value $\alpha$ is computed correctly.}
We know that $\alpha = \nest(G[T])$. If $s(P) = \emptyset$, then $G[T]$ is the disjoint union of
$G[T_1],\dots,G[T_p]$ with $\alpha_i = \nest(G[T_i])$, so $\alpha =
\max\{\alpha_1,\dots,\alpha_p\}$.  Otherwise, we get the situation from Fig.~\ref{fig:intuition}i.

First, we argue that $G[T]$ has a representation $\calR$ of nesting $\alpha$ from the formula.
Intervals of all subtrees except for the leftmost subtree $T_s$ and the rightmost one $T_t$ are
completely nested inside $\inter w$; and we minimize over all possible choices of $s \ne t$. Let
$\calR_i$ be a minimal representation for $G[T_i]$ from Lemma~\ref{lem:minimal_representation}, and
we use $\calR^\lftrt_s$ for $G[T_s]$. For every $i \ne s,t$, we get that $\nest(\calR_i) = \alpha_i$
is increased by one with $\inter w$. For $\calR_s$ and $\calR_t$, only $\nest^\lft(\calR_s^\lftrt) =
\beta_s$ and $\nest^\rt(\calR_t) = \beta_t$ are increased by one with $\inter w$. We get
\begin{linenomath*}
$$
\nest(\calR) = \nest(w) =
	\min_{s \ne t} \max \{\beta_s+1,\beta_t+1,\alpha_i+1 : i \ne s,t\} = \alpha.
$$
\end{linenomath*}

On the other hand, consider a representation $\calR$ of $G[T]$. There is no chain of nested
intervals containing intervals from two different subtrees $T_i$ and $T_j$. Let $\calR_i =
\calR[T_i]$ and let $\calR_s$ and $\calR_t$ be the leftmost and the rightmost of these
representations, respectively. For every $i \ne s,j$, the representation $\calR_i$ has the nesting
at least $\alpha_i$, so $\nest(\calR) \ge \alpha_i+1$. By Lemma~\ref{lem:meaning_of_gamma}, we know
that $\nest^\lft(\calR_s) \ge \beta_s$ and $\nest^\rt(\calR_t) \ge \beta_t$ and these chains are nested in
$\inter w$, so $\nest(\calR) \ge \max\{\beta_s+1,\beta_t+1\}$. So $\nest(\calR) \ge \alpha$ from the
formula.

\emph{The value $\beta$ is computed correctly.}
First, we construct a representation $\calR_\beta$ of $G_\beta$ with nesting $\beta+1$. If $s(P) = \{w\}$, in
every cleaned representation, $\inter w \not\subsetneq \inter{u_\beta}$, so that every other interval is
either nested in both, or in neither. So we can assume that $s(P) = \emptyset$.

When the added intervals are placed on the right of $\inter{u_\beta}$, intervals of all subtrees
except for a left-most one $T_s$ are completely nested inside $\inter{u_\beta}$; and we again
minimize over all possible choices of $s$.  Let $\calR_i$ be a minimal representation of $G[T_i]$,
we use $\calR_s^\lftrt$ for $G[T_s]$. For every $i \ne s$, we get that the nesting $\nest(\calR_i) =
\alpha_i$ is increased by one by $\inter{u_\beta}$. For $T_s$, the nesting
$\nest^\lft(\calR_s^\lftrt) = \beta_s$ is increased by one with $\inter{u_\beta}$. We get
\begin{linenomath*}
$$
\nest(\calR_\beta) = \nest(u_\beta) = \min_s \max \{\beta_s+1, \alpha_i+1 : i \ne s\} = \beta+1.
$$
\end{linenomath*}
For the other implication, consider a cleaned representation of $G_\beta$.  Similarly, as above, we
get that the nesting is at least $\beta+1$.

\emph{The value $\gamma$ is computed correctly.}
We just sketch the argument, it is similar as above. Let $\calR_i$ be a minimal representation of
$G[T_i]$. We may choose $T_s$ and $T_t$, and use $\calR_s^\lftrt$ for $T_s$.  Then only the nesting
$\nest^\lft(\calR_s^\lftrt) = \beta_s$ is increased by one with $\inter{v_\gamma}$ and only the
nesting $\nest^\rt(\calR_t) = \beta_t$ is increased by one with $\inter{u_\gamma}$. But since
$\calR_s^\lftrt$ is nested inside $\inter{u_\gamma}$ and $\calR_t$ is nested inside
$\inter{v_\gamma}$, it does not matter and the nestings $\nest(\calR_s^\lftrt)$ and $\nest(\calR_t)$
are both increased by one anyway. Therefore, this choice of $T_s$ and $T_t$ is useless and a
constructed representation of $G_\gamma$ has the nesting $\max\{\alpha_1,\dots,\alpha_p\}+1$. The
other implication is proved similarly as before.
\end{proof}

\begin{lemma} \label{lem:p_node_complexity}
For a P-node with $p$ children, the triple $(\alpha,\beta,\gamma)$ can be computed in $\O(p)$.
\end{lemma}

\begin{proof}
By Lemma~\ref{lem:triple_bounds}, we always have either $\beta_i = \alpha_i - 1$, or $\beta_i =
\alpha_i$. Only in the
former case, we may improve the nesting by choosing $s = i$ or $t = i$. We call subtrees $T_i$ with
$\beta_i = \alpha_i - 1$ as \emph{savable}.

For $\gamma$ and for $\alpha$ with $s(P) = \emptyset$, we just find the maximum $\alpha_i$ which can be done in
time $\O(p)$. For $\alpha$ with $s(P) \ne \emptyset$ and $\beta$, we first locate all $T_i$ which maximize
$\alpha_i$. If at least one of them is not savable, say $T_j$, then $\alpha = \alpha_j+1$ and
$\beta = \alpha_j$. Otherwise
if all are savable, then the values $\alpha$ and $\beta$ depend on the number of these subtrees. If there are
at most two, then $\alpha = \alpha_i$, otherwise $\alpha = \alpha_i+1$. If there is exactly one,
then $\beta = \beta_i$, otherwise $\beta = \beta_i+1$.
\end{proof}

\subsection{Triples for Q-nodes}

\begin{figure}[t!]
\centering
\includegraphics[width=\textwidth]{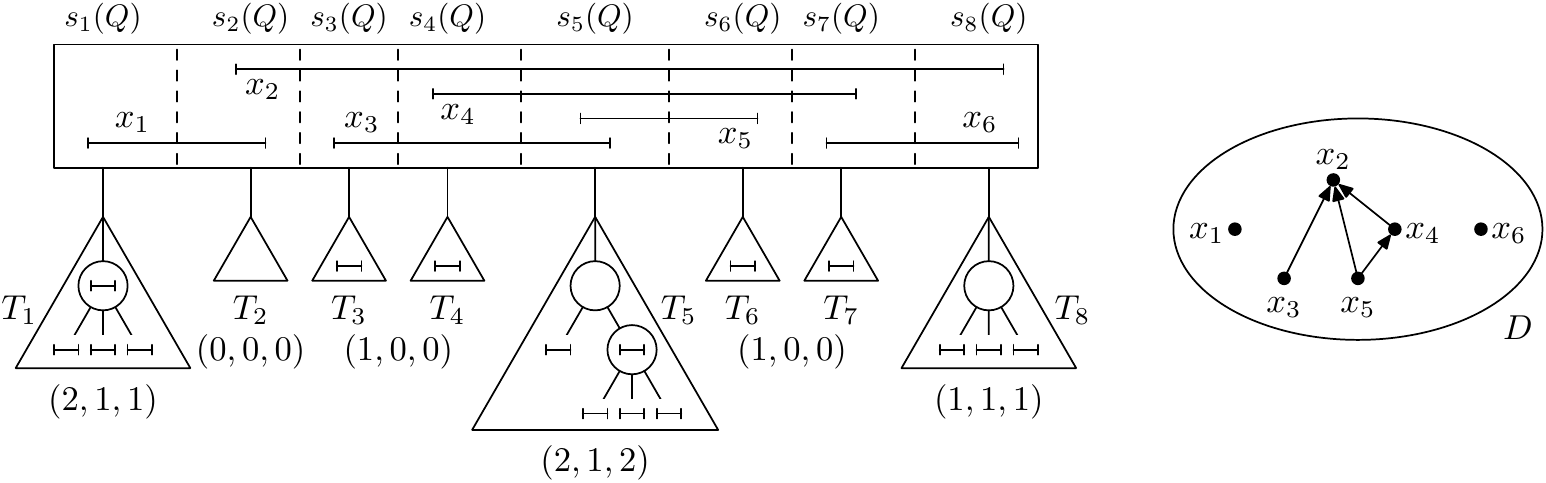}
\caption[]{On the left, a Q-node $Q$ with eight subtrees. On the right, the DAG $D$ of forced
nestings in $s(Q)$.\\
$\bullet$\quad For $\calR^*_5 = \calR_5$ (depicted), we get $\nest(x_3) = 2$, $\nest(x_5) = 3$,
$\nest(x_4) = 4$, and $\nest(x_2) = 5$.\\
$\bullet$\quad For $\calR^*_5 = \calR_5^\lftrt$ (by flipping $T_5$), we get $\nest(x_3) = 3$,
$\nest(x_5) = 2$, $\nest(x_4) = 3$, and $\nest(x_2) = 4$.\\
The second option minimizes the nesting and it gives the triple $(4,3,4)$ for $T[Q]$.}
\label{fig:q_node_example}
\end{figure}

The situation is more complex and the values $\gamma$ are also required. Let $Q$ be a Q-node with
subtrees $T_1,\dots,T_q$, where $q \ge 3$, each with a triple $(\alpha_i,\beta_i,\gamma_i)$. We want to compute
the triple $(\alpha,\beta,\gamma)$ of the subtree $T = T[Q]$.  Since lengths of chains are not changed by flipping
$Q$, we can fix the left-to-right order of its subtrees as $T_1,\dots,T_q$. See
Fig.~\ref{fig:q_node_example} for an example.

\heading{Structure of Chains.}
Suppose that we choose some cleaned representations $\calR_1,\dots,\calR_q$ of
$G[T_1],\dots,G[T_q]$. Then the corresponding cleaned representation of $G[T]$ is uniquely
determined. What is the structure of chains of nested intervals? Each chain starts in some subtree
$T_i$ and then continues with intervals in $s(Q)$ as follows. If it contains $\inter x \subsetneq
\inter y$ for $x,y \in s(Q)$, then $x \subsetneq_F y$, so $y$ starts more to the left and ends more
to the right than $x$ in sections of $Q$. We represent the relation $\subsetneq_F$ on $s(Q)$ by a
DAG $D$, having an edge $(x,y)$ if and only if $x \subsetneq_F y$; see Fig.~\ref{fig:q_node_example}
on the right.

Suppose that a chain of nested intervals of $s(Q)$ of lenght $\ell$ starts with $\inter x$. Let
$s_x^\lft(Q) = s_s(Q)$ and $s_x^\rt(Q) = s_t(Q)$, for some $s < t$. Then every chain of every
$\calR_i$ such that $s < i < t$ is nested in $\inter x$, so for each $\calR_i$, there exists a chain
of nested intervals of length $\nest(\calR_i)+\ell$. But there might not be chains of lengths
$\nest(\calR_s)+\ell$ and $\nest(\calR_t)+\ell$. The reason is that only chains in $\calR_s$ not
ending with an interval contained in the leftmost maximal clique of $\calR_s$ are nested in $\inter
x$, and only those of $\calR_t$ avoiding the rightmost maximal clique of $\calR_t$. So there only
exist chains of lengths $\nest^\rt(\calR_s)+\ell$ and $\nest^\lft(\calR_t)+\ell$.

By Lemma~\ref{lem:minimal_representation}, there exists a minimal representation $\calR_i$ of $G[T_i]$
with $\nest(\calR_i) = \alpha_i$, $\nest^\rt(\calR_i) = \beta_i$ and $\nest^\lft(\calR_i) = \gamma_i$; and we can
swap the last two nestings with $\calR_i^\lftrt$.
Let $\calR_i^* \in \{\calR_i,\calR_i^\lftrt\}$. We denote $\bigcirc_i^\rt = \nest^\rt(\calR_i^*)$
and $\bigcirc_i^\lft = \nest^\lft(\calR_i^*)$.
\begin{linenomath*}
\begin{equation} \label{eq:choices_Q_node}
\text{\emph{For each $T_i$, we choose either}}\qquad
\calR_i^* = \calR_i:\quad
\begin{matrix}
\bigcirc_i^\rt = \beta_i,\\
\bigcirc_i^\lft = \gamma_i,
\end{matrix}
\qquad\text{\emph{or}}\qquad
\calR_i^* = \calR_i^\lftrt:\quad
\begin{matrix}
\bigcirc_i^\rt = \gamma_i,\\
\bigcirc_i^\lft = \beta_i.
\end{matrix}
\end{equation}
\end{linenomath*}
By combining these chosen representations $\calR_i^*$ for all subtree $T_i$, we get one of $2^q$ possible
representations $\calR[T]$. For each of them, we compute $\nest(x)$ for all $x \in s(Q)$
using the following formulas:
\begin{linenomath*}
\begin{equation} \label{eq:nesting_in_Q_node}
\begin{split}
\nest(x) = \max \{\bigcirc_s^\lft+1, \bigcirc_t^\rt+1, \alpha_i+1, \nest(y)+1:
\qquad\qquad\qquad\qquad\qquad\\
	\text{$s_x^\lft(Q) = s_s(Q)$, $s_x^\rt(Q) = s_t(Q)$, $s < i < t$ and $y \in \pred_D(x)$}\bigr\},
\end{split}
\end{equation}
\end{linenomath*}
where $\pred_D(x)$ denotes the set of all direct predecessors of $x$ in $D$.  These values can be
computed according to a topological sort of $D$.

\heading{Formulas for Triples.}
The triple $(\alpha,\beta,\gamma)$ is determined by minimal nestings of $G_\alpha$, $G_\beta$, and
$G_\gamma$. We study how chains in $s(Q)$ are extended by the added intervals $\inter{u_\alpha}$,
$\inter{u_\beta}$, $\inter{u_\gamma}$ and $\inter{v_\gamma}$.  Further, we consider two copies
$\inter{u_{\beta^\lft}}$ and $\inter{u_{\beta^\rt}}$ of $\inter{u_\beta}$. Recall that the left-to-right
ordering of the subtrees of $Q$ is fixed.  Therefore, $\inter{u_\beta}$ can intersect $\calR$ either
from left (represented by $\inter{u_{\beta^\rt}}$), or from right (represented by
$\inter{u_{\beta^\lft}}$). Similarly, we assume that $\inter{u_\gamma}$ intersects $\calR$ from left
while $\inter{v_\gamma}$ from right.

We add auxiliary vertices $u_\alpha$, $u_\beta^\lft$, $u_\beta^\rt$, $u_\gamma$ and $v_\gamma$ into
$D$ and get the following extended DAG $D'$:
\begin{linenomath*}
\begin{eqnarray*}
V(D') &=& V(D) \cup \{u_\alpha, u_{\beta^\lft}, u_{\beta^\rt}, u_\gamma, v_\gamma\}\\
E(D') &=& E(D) \cup \bigl\{(x,u_\alpha),(y,u_{\beta^\lft}),(y,v_\gamma),(z,u_{\beta^\rt}),(z,u_\gamma) :\\
&&\qquad\qquad\qquad x \in s(Q), y \in s(Q) \setminus s_1(Q), z \in s(Q) \setminus s_q(Q)\bigr\}.
\end{eqnarray*}
\end{linenomath*}
In other words, $u_\alpha$ extends every chain in $s(Q)$, but $u_{\beta^\lft}$ and $v_\gamma$ extend
only those not ending with an interval in $s_1(Q)$, and $u_{\beta^\rt}$ and $u_\gamma$ only those not
ending with an interval in $s_q(Q)$.

We compute $(\alpha,\beta,\gamma)$ of $T[Q]$ using the following formulas:
\begin{linenomath*}
\begin{eqnarray*}
\alpha &=& \min_{\forall \calR_i^*}
	\max \bigl\{\alpha_1,\dots,\alpha_q, \nest(y) : y \in \pred_{D'}(u_\alpha)\bigr\},\\
\beta^\lft &=& \min_{\forall \calR_i^*}
	\max \bigl\{\beta_1,\alpha_2,\dots,\alpha_q, \nest(y) : \text{$y \in \pred_{D'}(u_{\beta^\lft})$}\bigr\},\\
\beta^\rt &=& \min_{\forall \calR_i^*}
	\max \bigl\{\alpha_1,\dots,\alpha_{q-1},\beta_q, \nest(y) : \text{$y \in \pred_{D'}(u_{\beta^\rt})$}\bigr\},\\
\beta &=& \min \{\beta^\lft, \beta^\rt\},\\
\gamma &=& \min_{\forall \calR_i^*}
	\max \bigl\{\alpha_1,\dots,\alpha_q, \nest(y) :
		\text{$y \in \pred_{D'}(u_\gamma) \cup \pred_{D'}(v_\gamma)$}\bigr\}.
\end{eqnarray*}
\end{linenomath*}

\begin{lemma} \label{lem:q_node_formulas}
The formulas compute the triple $(\alpha,\beta,\gamma)$ of $T[Q]$ correctly.
\end{lemma}

\begin{proof}
We assume that the left-to-right order of subtrees of $Q$ is fixed, it does not change nesting.
Recall that in a cleaned representation $\calR$ of $G[T]$, the nesting $\nest(\calR)$ is determined
by representations $\calR[T_1],\dots,\calR[T_q]$.

\emph{The value $\alpha$ is computed correctly.}
First, we construct a representation $\calR_\alpha$ of $G_\alpha$ with $\nest(\calR_\alpha) =
\nest(u_\alpha) = \alpha+1$ for $\alpha$ given by the above formula. We construct $2^q$
representations for all choices of $\calR_i^*$ using (\ref{eq:choices_Q_node}), and we use a
representation minimizing the nesting, corresponding to the minimum $\min_{\forall \calR_i^*}$ in
the formula.  The choices $\calR_i^*$ determine a cleaned representation $\calR_\alpha$ of
$G_\alpha$.  Nesting of the intervals of $s(Q)$ is computed using (\ref{eq:nesting_in_Q_node}) and
$\nest(u_\alpha)$ is equal the length of the longest chain in $\calR_\alpha[G[Q]]$ increased by one.
The formula for $\alpha$ maximizes over lengths of all chains in $\calR_\alpha[G[Q]]$.

On the other hand, consider a cleaned representation $\calR$ of $G[T]$. We argue that $\nest(\calR)
\ge \alpha$ for $\alpha$ given by the above formula. By Lemma~\ref{lem:meaning_of_gamma}, the
representation $\calR[T_i]$ has $\nest(\calR[T_i]) \ge \alpha_i$ and either $\nest^\rt(\calR[T_i])
\ge \beta_i$ and $\nest^\lft(\calR[T_i]) \ge \gamma_i$, or $\nest^\rt(\calR[T_i]) \ge \gamma_i$ and
$\nest^\lft(\calR[T_i]) \ge \beta_i$. As in the proof of Lemma~\ref{lem:local_minimality}, by
replacing $\calR[T_i]$ with $\calR_i$ in the former case and with $\calR_i^\lftrt$ in the latter
case, we do not increase the nesting. We obtain a representation of $G[T]$ of nesting $\alpha$, so
$\nest(\calR) \ge \alpha$.

\emph{The value $\beta$ is computed correctly.}
Concerning $\beta$, in every cleaned representation $\calR_\beta$ of $G_\beta$, either intervals of
$s_q(Q)$ are not nested in $\inter{u_\beta}$ (represented by $\inter{u_{\beta^\lft}}$), or intervals
of $s_1(Q)$ are not nested in $\inter{u_\beta}$ (represented by $\inter{u_{\beta^\rt}}$). We compute
both possibilities in $\beta^\lft$ and $\beta^\rt$, and use the minimum. The rest of the arguments
is similar as above.

\emph{The value $\gamma$ is computed correctly.}
Again, the arguments are similar as for $\alpha$ above, the only difference is that $\inter{u_\alpha}$ is replaced by
both $\inter{u_\gamma}$ and $\inter{v_\gamma}$.
\end{proof}

Unfortunately, formulas do not directly lead to a polynomial-time algorithm since they minimize
over $2^q$ possible choices of $\calR_i^*$. Next, we prove that these
choices can be done greedily.

\begin{lemma} \label{lem:q_node_greedy}
For each of $\alpha$, $\beta^\lft$, $\beta^\rt$ and $\gamma$, we can locally choose $\calR_i^*$
minimizing the value.
\end{lemma}

\begin{proof}
Notice that the choices of $\calR_i^*$ are independent of each other since each $\calR_i^*$
influences only lenghts of chains starting in $\calR[T_i]$.  We give a description for $\alpha$, and it
works similarly for the others.
	
For each $x \in s(Q)$, we compute the length $\ell(x)$ of a longest chain in $s(Q) \cup
\{u_\alpha\}$ starting with $\inter x$. Let
\begin{linenomath*}
$$
\ell_i^\lft = \max_{\substack{x \in s(Q)\\ s_x^\lft(Q) = s_i(Q)}} \ell(x),\qquad\text{and}\qquad
\ell_i^\rt = \max_{\substack{x \in s(Q)\\ s_x^\rt(Q) = s_i(Q)}} \ell(x),
$$
\end{linenomath*}
and let $\ell_i^* = 0$ if no such $x \in s(Q)$ exists. We choose $\calR_i^* = \calR_i$ if and only
if $\ell_i^\rt \ge \ell_i^\lft$. These choices minimize lengths of all chains in a representation of
$G[T]$. For instance, in Fig.~\ref{fig:q_node_example}, we get $\ell_5^\lft = 3$ and $\ell_5^\rt =
2$, so we choose $\calR_5^* = \calR_5^\lftrt$. 
\end{proof}

\begin{lemma} \label{lem:q_node_complexity}
For a Q-node $Q$ with $q$ children, the triple $(\alpha,\beta,\gamma)$ of $T[Q]$ can be computed in time
$\O(q+m_Q)$, where $m_Q$ is the number of edges of $G[s(Q)]$.
\end{lemma}

\begin{proof}
Since $G[s(Q)]$ is connected, it contains at most $m_Q$ vertices. For every $x \in s(Q)$,
we know $s_x^\lft(Q)$ and $s_x^\rt(Q)$ which we use to compute the DAG $D$.  This can be done by
considering all $m_Q$ edges, and testing for each whether the pair is nested.  Then, we construct
the extended DAG $D'$.

For each $x \in V(D')$ and each $y \in \{u_\alpha,u_\beta^\lft,u_\beta^\rt,u_\gamma,v_\gamma\}$, we
compute the length of a longest path from $x$ to $y$. This can be done in linear time for all
vertices $x$ by processing $D'$ from the top to the bottom. For each $T_i$, we choose greedily
$\calR_i^*$ as described in the proof of Lemma~\ref{lem:q_node_greedy}. We compute the triple
$(\alpha,\beta,\gamma)$ using the above formulas.  The total running time is $\O(q+m_Q)$.
\end{proof}

\subsection{Construction of Linear-time Algorithm}

We use the above results to prove that $\nest(G)$ can be computed in linear time:

\begin{proof}[Theorem~\ref{thm:recog_nint}]
For an interval graph $G$, we compute its MPQ-tree in time $\O(n+m)$~\cite{korte_mohring}. Then we
process the tree from the leaves to the root and compute triples $(\alpha,\beta,\gamma)$ for every
node, as described above. We output $\alpha$ of the root which is the minimal nesting number
$\nest(G)$. By Lemmas~\ref{lem:p_node_formulas} and~\ref{lem:q_node_formulas}, this value is
computed correctly.  By Lemmas~\ref{lem:p_node_complexity} and~\ref{lem:q_node_complexity}, the
running time of the algorithm is $\O(n+m)$.
\end{proof}

\section{Conclusions} \label{sec:nest_len_int_conclusions}

In this paper, we have introduced $k$-nested interval graphs which is a new hierarchy of graph
classes between proper interval graphs and interval graphs. The presented understanding is already
much greater than understanding of $k$-length interval graphs reached after more than 35 years of their
research. We have presented a relatively simple recognition algorithm based on dynamic programming
and minimal representations. Other research directions immediately open. In~\cite{hkko}, our results
are used to derive minimal forbidden induced subgraphs of $\nint{k}$.

\begin{problem}
Which structural properties and characterizations of proper interval graphs generalize to $k$-nested
interval graphs?
\end{problem}

\begin{problem}
Which computational problems solvable efficiencly for proper interval graphs can be solved
efficiently for $k$-nested interval graphs as well?
\end{problem}

The second problem is interesting for computational problems which are harder for general interval
graphs. One example is deciding first-order logic properties which is \cW2-hard for interval
graphs~\cite{fo_checking}, but can be solved in \cFPT for $k$-nested interval
graphs~\cite{fo_posets}.

\begin{figure}[t!]
\centering
\includegraphics{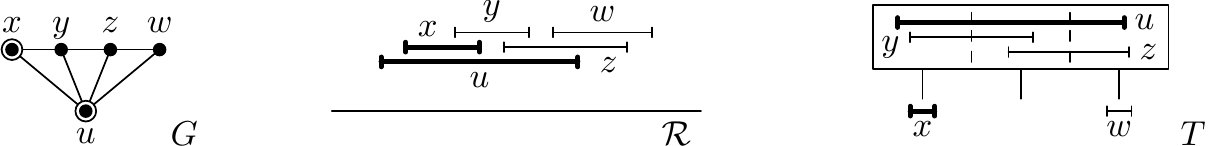}
\caption{On the left, a proper interval graph $G$, i.e., $\nest(G) = 1$. In the middle, an example of
an extending representation $\calR$ with $\nest(\calR) = 2$, and its nesting is optimal since
$\inter x' \subsetneq \inter u'$. Further, in every extending representation, $\inter x' \subsetneq
\inter y$ or $\inter y \subsetneq \inter u'$. On the right, the corresponding MPQ-tree $T$.}
\label{fig:non_extendible_pint}
\end{figure}

In~\cite{kos17b}, our results are used to attack the problem $\ext(\nint{k})$. A polynomial-time
algorithm for finding an extending interval representation of minimal nesting in derived, by a much
more involved dynamic programming than in the recognition algorithm of
Section~\ref{sec:recognizing_nesting}.  A partial representation $\calR'$ poses three restrictions:
\begin{packed_itemize}
\item[(i)] Some pre-drawn intervals can be nested in each other which increases the nesting.
\item[(ii)] The consecutive ordering has to extend $\wlt$ which restricts the possible shuffling of subtrees.
\item[(iii)] Some subtrees can be optimized differently depending on the side they are attached.
\end{packed_itemize}
The problem is difficult since we have to deal with them simultaneously. For an example, see
Fig.~\ref{fig:non_extendible_pint}.

\heading{Acknowledgment.} We want to thank Takehiro Ito and Hirotaka Ono for fruitful discussions.

\bibliographystyle{plain}
\bibliography{nested_and_length_limited_int}

\end{document}